\magnification=1200 
\parskip = 10pt plus 5pt 
\parindent = 15pt 
\baselineskip = 15pt 
\input amssym.def 
\input amssym
\input epsf.tex
\line{\hfil{Report 98-34~~~~~~~~~}} 
\line{\hfil{Revised~version~~~~~~}}
\line{\hfil{December, 1999~~~~~~}}
\topglue 0.95in 
\pageno = 0
\footline={\ifnum \pageno < 1 \else \hss \folio \hss \fi} 
\centerline{\bf SPIN STRUCTURES ON RIEMANN SURFACES}
\centerline{\bf AND THE PERFECT NUMBERS}
\vskip .65in 
\centerline{\bf Simon Davis}
\vskip .5in 
\centerline{School of Mathematics and Statistics}
\vskip 1pt
\centerline{University of Sydney}
\vskip 1pt
\centerline{NSW 2006, Australia}
\vskip 25pt 
\noindent{\bf Abstract.}  The compositeness of Mersenne numbers can be
viewed in terms of equality with sums of consecutive integers.
This can be conveniently described by partitioning an array of sites  
representing the Mersenne number.  The sequence of even perfect numbers also
can be embedded in a sequence of integers, each equal to the number of 
odd spin structures on a Riemann surface of given genus.  A condition for
the existence of odd perfect numbers is given in terms of the rationality 
of the square root of a product containing, in particular, a sequence of 
repunits.  It is shown that rationality of the square root expression 
depends on the characteristics of divisors of the repunits.

\vfill \eject

\noindent
{\bf 1. Introduction}

The geometry of superstring perturbation theory is based on the properties of
moduli spaces and spin structures on Riemann surfaces.  Superstring scattering
amplitudes can be represented by integrals over supermoduli space, a
Grassmann manifold with $3g-3$ even parameters and $2g-2$ odd parameters,
which can be reduced to integrals over a ramified covering of moduli space 
with each copy corresponding to a different spin structure.  The action in 
the weighting factor contains both bosonic and fermionic fields, and  
even and odd spin structures are distinguished by the change
in sign when the world-sheet coordinate of a fermion field traverses an 
A-cycle or B-cycle;  of the $2^{2g}$ spin structures obtained by assigning 
$+$ or $-$ signs to each of the $2g$ cycles, there are $2^{g-1}(2^g+1)$
even spin structures and $2^{g-1}(2^g-1)$ odd spin structures.  
The moduli space at any genus splits
into two components, even spin moduli space $M_g^+$ and odd spin moduli space 
$M_g^-$, and the modular group acts on both components separately.  The 
equivalence of the number of odd spin structures on a genus-$g$ Riemann 
surface and the perfect numbers, when $2^g-1$ is a Mersenne prime,
suggests that a geometrical representation of the sequence of Mersenne primes
may provide information about the properties of perfect numbers.

In this paper, it is shown that the geometrical representation of
Mersenne numbers leads to a criterion for selecting those numbers
which are primes.  This is based on a theorem which makes use of 
a common property of the decomposition of both factors $2^{g-1}$ and
$2^g-1$, which is related to the binary system underlying
the equivalence between the counting of the total number of spin structures
at genus $g$ and the corresponding perfect number.  

The existence of odd perfect numbers is shown to be related to the 
irrationality of the square root of a product including quotients of the
form ${{x^n-1}\over {x-1}}$, repunits in the base $x$.   Although there are 
only a few cases of quotients of this type being equal or perfect squares, 
their products might be squares of integers, as one can verify when $n~=~3$.  
For example,  the existence of integer solutions to a quadratic Diophantine 
equation is used to show that 
${{x_1^3-1}\over {x_1-1}} {{x_2^3-1}\over {x_2-1}}$ can be the square 
of an integer.

There has been considerable amount of research on divisors of Lucas and
Lehmer sequences, which include the Fibonacci and Pell sequences as
particular examples.  The repunits ${{x^n~-~1}\over {x~-~1}}$ form a special
Lucas sequence and their divisors include the cyclotomic polynomials
$\Phi_d(x)$, $d\vert n$.  A study of the monotonicity of these polynomials
for $x\ge 1$, and their divisors, provides an indication of
whether the entire product of repunits contains unmatched prime divisors
with irrational square roots.

\noindent
{\bf 2. The Geometrical Representation of Mersenne Numbers and Spin}
\hfil\break 
\phantom{.....}{\bf Structures on Riemann Surfaces}

The existence of the even perfect numbers $2^{n-1}(2^n~-~1)$ is contingent 
upon the primality of the Mersenne number $M_n=2^n-1$ [1][2], and
the Lucas-Lehmer test [3][4][5] is known to be satisfied by 
38 Mersenne primes.  The prime divisors of $M_n$ must have the form 
$2nk+1$ [6] and $8k \pm 1$ [7][8] and
additional primality tests for $N= 2^n - 1$ have been developed using 
the factorization of $N - 1$ [9][10].

Both even perfect numbers [2][11], and Mersenne primes [12], lend themselves 
to geometrical interpretation.  Even perfect numbers are known to 
be triangular and hexagonal [11], and  it has been shown for example that
circular planar nearrings based on $({\Bbb Z}_q,+,*)$, where $*$ is a 
multiplication required for planar nearrings, have special properties
when $q$ is a Mersenne prime [12].

The Mersenne numbers $2^n - 1$ can be 
represented in a triangular array for finite $n$ by placing $2^m$ sites at 
the $m^{th}$ level.  If the angle at the apex of the triangle 
is fixed, the base increases linearly with the level
number and has length $l_m~=~m l_1$ where $l_1$ is the distance between the
two sites at the $m=1$ level.   The distance between neighbouring sites
at the $m^{th}$ level is ${{l_m}\over {2^m~-~1}}~=~{m\over {2^m~-~1}}~l_1$.
Bisection of this triangle and inclusion of the apex in one of the sets is
a geometrical representation of $2^n - 1$ as the sum of two consecutive 
numbers $2^{n-1} - 1$ and $2^{n-1}$.  Division of the triangle into more
than two approximately equal parts is related to the following lemma [13]
[14]:
\vskip 10pt
\noindent
{\bf Lemma} (de la Rosa 1978) -  A positive integer is a prime or a power of 
$2$ if and only if it cannot be expressed as the sum of at least three 
consecutive positive integers. 

Suppose that the triangle is divided into $K$ parts. Then the site
located at a fraction of the distance along the $m^{th}$ level, 
${{\bar m}\over {2^m-1}}~\cdot~l_m$,
will be included in the $j^{th}$ triangle if 
$${{(j~-~1)(2^m~-~1)}\over K}~\le~{\bar m}~\le~{{j(2^m~-~1)}\over K}
\eqno(1) 
$$
The number of sites included in the $j^{th}$ triangle is
$$N_m^K~=~\left[{{j(2^m~-~1)}\over K}\right]~-~\left\{{{(j-1)(2^m~-~1)}
\over K}\right\}~+~1
\eqno(2)
$$
\vskip 10pt

\vbox{
\epsfysize=3in
\epsfxsize=3.5in
\centerline{\epsfbox{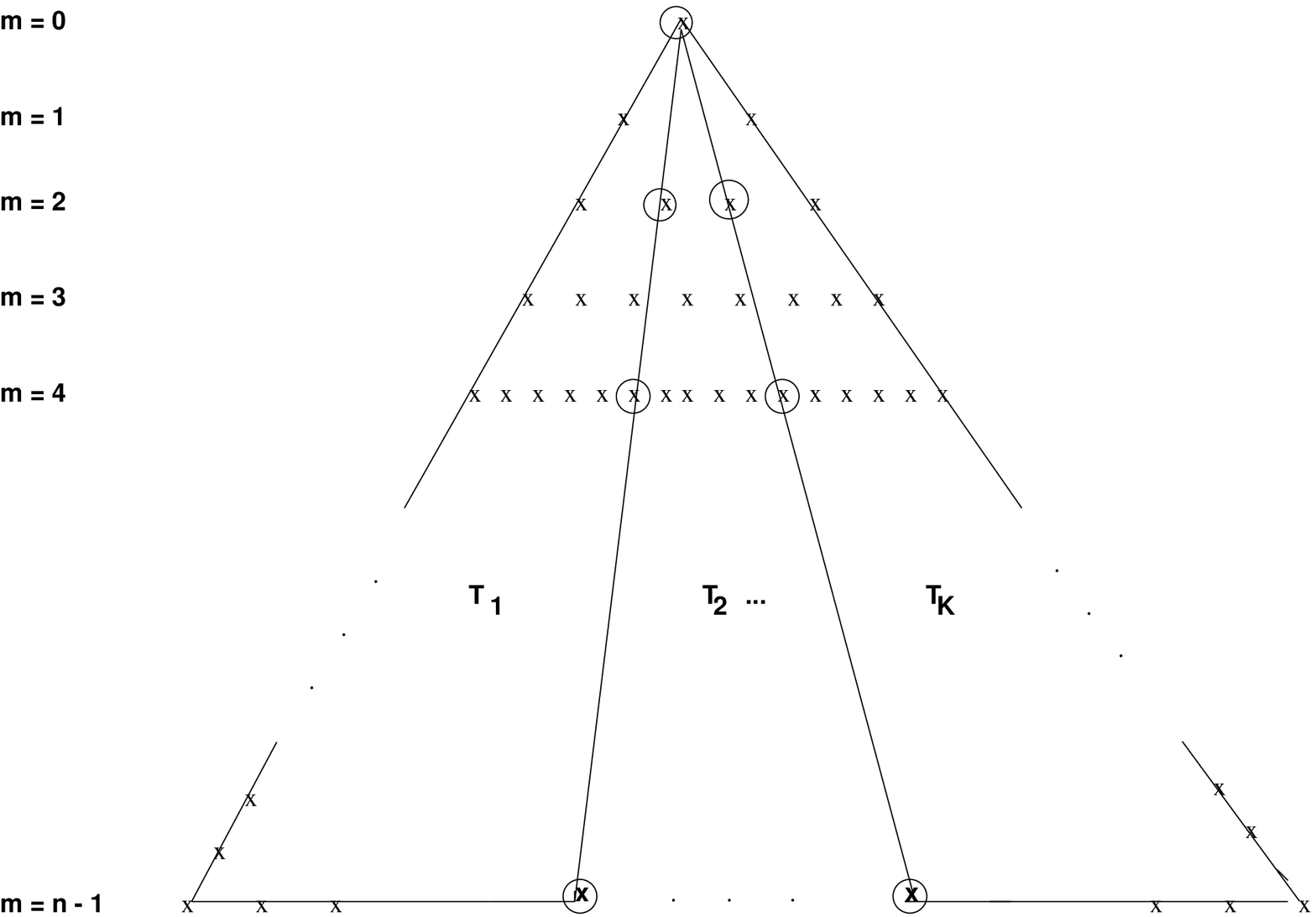}}
\vskip 0.2in
\noindent{\bf Fig. 1.  Triangular Representation of the Mersenne Number 
$2^n~-~1$.}
\hfil\break
\phantom{..............}{\bf The circled sites are shared by adjacent 
triangles.}}

If the original triangle is divided into $K$ parts so that
$N_m^K~=~{{2^m~-~1}\over K}~+~1$ when $K\vert 2^m~-~1$, $m~>~1,~m~\vert~n-1$,
the total number of sites in the triangle $T_j$, including the apex, is
$$1~+~\sum_{m=1}^{n-1}~N_m^K~\doteq~1~+~\sum_{m\atop {K\vert {2^m-1}}}~
\left({{2^m~-~1}\over K}~+~1\right)~+~\sum_{m\atop {K~\not\vert~{2^m-1}}}~
\left(\left[{{2^m~-~1}\over K}\right]~+~1\right)
\eqno(3)
$$
Let $\tau_2(n~-~1,~K)$ be the number of values of $m$ such that $m~=~0$,
for the apex of the triangle, or $m~\vert~n~-~1$ and $K~\vert~2^m~-~1$; 
it satisfies the inequality $\tau_2(n~-~1,K)~\le~\tau(n~-~1)$.  
The number of sites in an interior triangle $T_j,~j~=~2,...,K-1$ 
would be approximately
$$\eqalign{1~+~{{2^{n-1}b(n-1)~-~1}\over K}~&+~\left(1~-~{1\over K}\right)
\tau_2(n-1,K)~+~\left[{{2^{n-1}(2~-~b(n-1))~-~3}\over K} \right]
\cr
~&+~\left[\left({1\over 2}~-~{1\over K}\right)(n~-~1~-~\tau_2(n-1,~K))
\right]
\cr}
\eqno(4)
$$
where $1~<~ b(n-1)~<~2$.
The number of shared sites in an interior triangle, including the
apex, is $2\tau_2(n~-~1,K)~-~1$, whereas the number of shared sites in the
outer triangles $T_1$ and $T_K$ is $\tau_2(n-1,K)$, and the overcounting of 
the sites in all of the triangles is given by $(K~-~1)\tau_2(n~-~1,K)$.   

If $P_g$ denotes the number $2^{g-1}(2^g - 1)$, then $P_g=4P_{g-1}+2^{g-1}$
follows from the decomposition of the Mersenne number $M_g$ into $M_{g-1}$
and $M_{g-1}+1$.  When $M_g$ is a Mersenne prime, this is the only
possible expression in terms of a sum of consecutive integers.  If a prime
other than the factor $2^g - 1$ is used, the recursion relation above 
would no longer be valid.  Similarly, composite integers other than $2^{g-1}$  
can be decomposed into three or more addends, which implies that there
are additional factors giving rise to a sum-of-divisors function $\sigma(N)$
not satisfying ${{\sigma(N)}\over N}~=~2$.

The factor of $4$ in the recurrence relation for $P_g$ is a reflection
of the equivalence between these integers and the number of odd spin
structures on a genus-$g$ Riemann surface.  
The counting of odd spin structures on a Riemann surface is based on a 
binary system, because the number of Dirac zero modes is either $0$ or 
$1$ mod 2 and it is additive when surfaces of genus $g_1$ and $g_2$ are 
joined.  Since there are one odd and three even spin structures on a 
torus, the odd spin structures at genus $g-1$ can be combined with any
of the three even spin structures at genus 1, and the even spin
structures at genus $g-1$ can be combined with the odd spin structure
at genus 1 to produce odd spin structures at genus $g$.  
This reveals the singlet-triplet structure underlying the
binary system and the number of ways of combining odd and 
even spin structures for each handle to produce an overall 
odd spin structure is
$$\eqalign{1~+~\left({g\atop 2}\right)~\cdot~3^2
~+~\left({g\atop 4}\right)~\cdot~3^4
~+~...~+~\left({g\atop {g~-~1}}\right)~\cdot~3^{g~-~1}~&=~
{{(1~+~3)^g~+~(1~-~3)^g}\over 2}
\cr
~&=~2^{g-1}(2^g~-~1)
\cr}
\eqno(5)
$$ 
when $g$ is odd.  The properties of the set of odd spin structures, when
$2^g-1$ is a Mersenne prime, which might continue indefinitely, shall be
described subsequently.

A spin structure ${\cal S}_\xi$ on $\Sigma$, which is a holomorphic line 
bundle ${\cal L}$ such that ${\cal L}^{\otimes 2}~=~{\cal K}$, the cotangent 
bundle, may also be viewed as a quadratic refinement 
$q_\xi:H_1(\Sigma,{\Bbb Z}_2)~\to {\Bbb Z}_2$, of an intersection form  
$\sigma(mod~2):H_1(\Sigma,{\Bbb Z}_2)\otimes H_1(\Sigma, {\Bbb Z}_2)
~\to~{\Bbb Z}_2$ [15] satisfying the property
$q_\xi(t_1~+~t_2)~=~q_\xi(t_1)~+~q_\xi(t_2)~+~\sigma(mod~2)(t_1,t_2),
~~~t_1,t_2~\in~{\Bbb Z}_2$.  The Arf invariant, which is zero if $q_\xi=0$
for more than half of the elements of $H_1(\Sigma, {\Bbb Z}_2)$ and $1$
otherwise, equals the parity of the theta characteristic
$\xi~=~\xi_1^\prime \xi_1^{\prime\prime}~+~...~+~\xi_g^\prime 
\xi_g^{\prime\prime}~(mod~2)$; it is also equivalent to the Atiyah invariant, 
the dimension, mod 2, of the holomorphic line bundle defined by the spin
structure on the surface $\Sigma$, which is zero $2^{g-1}(2^g + 1)$ times
and equal to one $2^{g-1}(2^g - 1)$ times [16][17].  Defining $\Gamma_g^+$
to be the subgroup of the mapping class group $\Gamma_g$ which leaves
invariant a quadratic refinement $q_\xi$ corresponding to an even spin
structure ${\cal S}_\xi$ and an even theta characteristic $\xi$, 
the even spin moduli space is $M_g^+~=~T_g/\Gamma_g^+$.  Similarly, if 
$\Gamma_g^-$ is the subgroup of the mapping class group which leaves
invariant an odd spin structure, then $M_g^-~=~T_g/\Gamma_g^-$ is the
odd spin moduli space.

At genus $g$, all odd spin structures can be generated by the application of
modular transformations to a set of $2^{g-1}$ spin structures.  The Ramond
sector $R$ consists of $2^g$ spin structures with genus-one components that 
are either even $(+-)$ or odd $(++)$.  By adding the genus-one components
and computing the overall parity of the genus-$g$ spin structure, it can 
be deduced that there will be $2^{g-1}$ even spin structures and $2^{g-1}$ 
odd spin structures in the Ramond sector.  At genus one, the modular group
$SL(2;{\Bbb Z})$, generated by $\tau~\to~\tau + 1$ 
and $\tau~\to~-{1\over \tau}$, where $\tau$ is the period of the torus,
interchanges the even spin structures $\{(+-), (-+), (--)\}$ and leaves
invariant the odd spin structure $(++)$.  One method for generating the
remaining odd spin structure is the application of products of genus-one 
transformations, acting on different handles, to the subset of odd spin 
structures $R_o$ in the Ramond sector.  Denoting the modular transformations
by $\rho_r, r=1,...,3^g-2^g-1$, it follows that $R_o~\cup~\cup_r~\rho_r(R_o)$
contains all of the odd spin structures at genus $g$.  

However, this technique does not entail the use of a minimal number of
transformations for generating these spin structures.
First, the genus-$g$ spin structure $(++++...++)$ is left invariant by all 
$\rho_r$ and therefore it appears in every set $\rho_r(R_o)$.  Secondly,
a genus-one modular transformation acting on only one handle 
alters $2^{g-2}$ spin structures in $R_o$, while a product of genus-one
modular transfomations acting on ${\ell}$ handles alters 
$2^{g-2}+2^{g-3}+...+2^{g-{\ell}} = 2^{g-{\ell}}(2^{\ell-1}-1)$ spin 
structures.  Since $2^{g-{\ell}}$ spin structures are unchanged, many of the
spin structures are counted repeatedly in the union
$R_o~\cup~\cup_r~\rho_r(R_o)$.  The presence of a fixed spin structure
$(++++...++)$ indicates that these modular transformations $\rho_r$ 
belong to the group $\Gamma_g^-$, which leaves invariant an odd spin 
structure ${\cal S}_\xi$.  

Since there are other modular transformations, belonging to the group 
$\Gamma_g^+$, which alter all of the spin structures in $R_o$, they may
be used to generate the odd spin structures with minimal overlap
between the different sets.  If there exist modular transformations
which induce no overlap, they may be denoted by $\sigma_r,~r=1,...,2^g-2$, 
and all odd spin structures would be included in the set 
$R_o~\cup~\cup_r~\sigma_r(R_o)$.

With an appropriate definition of the action of $\sigma_r$ on the remaining
odd spin structures, the set $\{{\bf 1},\sigma_r\}$ can be mapped 
isomorphically onto the multiplicative group $G_g$ of non-zero elements 
of a finite field $({\Bbb Z}_{2^g-1},\cdot,+)$ when $2^g-1$ is prime.  The 
order of $G_g$ is $\vert G_g\vert~=~2^g-1$ and the group does not have 
any proper subgroups.  It is an hereditary field group, which has the 
property that any subgroup, if it exists, would be a field group [17].  
Moreover, whenever $G$ is an hereditary field group and $\vert G\vert$ 
is odd, either $\vert G\vert = 1$ or $\vert G\vert$ is a Mersenne prime 
[18].  When $2^g-1$ is not a Mersenne prime, one may expect that
there will be subgroups of $G_g$ which are not field groups.

\noindent
{\bf 3. Odd Perfect Numbers}

Lower bounds of $10^{300}$ for odd perfect numbers [19] and
$10^6$ for the largest prime factor [20] of an odd perfect
number have been obtained.  No odd numbers have been found to satisfy the
condition ${{\sigma(N)}\over N}~=~2$, although there are odd integers, 
with five distinct prime factors, which produce a ratio nearly equal to 2:
$2~-~10^{-12}~<~{{\sigma(N)}\over N}~<~2~+~10^{-12}$ [21].
It has been established that any odd perfect number should take the form
$$N~=~(4k~+~1)^{4m+1}~s^2
\eqno(6)
$$
where $4k+1$ is a prime number with the property $gcd(4k+1,s)~=~1$ [22].
Using the prime decomposition of $s$, it follows that
$$s^2~=~q_1^{2\alpha_1}~...~q_l^{2\alpha_l}
\eqno(7)
$$
and 
$$\sigma(s^2)~=~\sigma(q_1^{2\alpha_1})~...~\sigma(q_l^{2\alpha_l})
             ~=~\prod_{i=1}^l~{{q_i^{2\alpha_i+1}~-~1}\over {q_i-1}}
\eqno(8)
$$
The ratio ${{\sigma(N)}\over N}$ equals
$$\left[{{(4k+1)^{4m+2}~-~1}\over {4k~(4k+1)^{4m+1}}}\right]~
{{\sigma(s^2)}\over {s^2}}~=~\left[{{(4k+1)^{4m+2}~-~1}\over
 {4k~(4k+1)^{4m+1}}}
\right]~\left[{{\sigma(s)^2}\over {s^2}}\right]~\left[{{\sigma(s^2)}
\over {\sigma(s)^2}}\right]
\eqno(9)
$$

The condition for $N$ to be a perfect number is ${{\sigma(N)}\over N}~=~2$
and it follows from equation (9) that
$${{\sigma(s)}\over s}~=~{\sqrt 2}~\prod_{i=1}^l~{{(q_i^{\alpha+1}-1)}
\over {(q_i~-~1)^{1\over 2}(q_i^{2\alpha_i+1}~-~1)^{1\over 2}} }
~\times~\left[{{4k(4k+1)^{4m+1}}\over {(4k+1)^{4m+2}~-~1}}\right]^{1\over 2}
\eqno(10)
$$ 
and
$$\prod_{i=1}^l~{1\over {(q_i^{\alpha_i+1}~-~1)}}~{{\sigma(s)}\over s}
~=~{\sqrt 2}~\prod_{i=1}^l~{1\over {(q_i^{2\alpha_i+1}~-~1)^{1\over 2}
(q_i~-~1)^{1\over 2} }}~\times~ \left[{{4k(4k+1)^{4m+1}}\over {(4k+1)^{4m+2}
~-~1}}\right]^{1\over 2}
\eqno(11)
$$
Since the product on the left-hand side is a rational number, the 
consistency of equation (11) depends on the rationality of the expression on 
the right-hand side of the equation.

The following theorem [23][24][25] may be used to determine whether 
$\left[{{(4k+1)^{4m+2}~-~1}\over {4k}}\right]$ is a square.

\noindent
{\bf Theorem} (Nagell 1921, Ljunggren 1943) - The integer solutions to the
equation 
$${{x^n~-~1}\over {x~-~1}}~=~y^2
\eqno(12)
$$
include
$$\eqalign{n~&=~2,~x~=~y^2~-~1,~y~\in~{\Bbb Z}
\cr
&~~~~~~~~~~~~~~if~x~is~prime,~then~x~=~3,~y~=~\pm 2
\cr
n~&=~3,~x~=~0,~y~=~\pm 1;~x~=~-1,~y~=~\pm 1
\cr
n~&=~4,~x~=~7,~y~=~20
\cr
n~&=~5,~x~=~3,~y~=~11
\cr}
\eqno(13)
$$
There are no primes $4k+1$ and integers $n=4m+2$ in this list such 
that
$\left[{{(4k+1)^{4m+2}~-~1}\over {4k}}\right]^{1\over 2}$ is a rational number.
Similarly, 
$$\prod_{i=1}^l~{1\over {(q_i^{2\alpha_i+1}~-~1)^{1\over 2}}}{1\over 
{(q_i-1)^{1\over 2}}}
~=~\prod_{i=1}^l~{1\over {(q_i^{2\alpha_i+1}-1)}}~[ 1~+~q_i~+~q_i^2~+~...
~q_i^{2\alpha_i}~]^{1\over 2}
\eqno(14)
$$ 
From the previous theorem, the number 
$[1~+~q_i~+~q_i^2~+~...~q_i^{2\alpha_i}]^{1\over 2}$ is rational when
$q_i~=~3,~\alpha_i~=~2$.   If
$3$ is a prime factor of $s$, this product can be expressed as

$$\left({{11}\over {242}}\right)^{\delta_{q_i,3}\delta_{\alpha_i,2}}
~\times~\prod_{{i=1}\atop {{q_i\ne 3}\atop {\alpha_i\ne 2}}}^l
~{1\over {(q_i^{2\alpha_i+1}-1)} }~
[1~+~q_i~+~q_i^2~+~...~q_i^{2\alpha_i}]^{1\over 2}
\eqno(15)
$$

Integer solutions to the equation
$$a{{x^n-1}\over {x-1}}~=~y^m
\eqno(16)
$$
have been listed for a restricted set of values of $a$ and $x$ [26].
\vskip 2pt
\noindent
{\bf Theorem} (Inkeri 1972).  The only solution of the equation
$$a{{x^n~-~1}\over {x-1}}~=~y^m~~~~~~(1~<~a~<~x~\le~10,~ n~>~2,~m~\ge~2)
$$
is $a=4$, $n=4$, $x=7$, $m=2$, $y=40$. If $10~<~x~<~15$, there are solutions,
if any, only in the cases $x=11$, $a=5, 7$ and $x=14$, $a=11$.

When $n=3$, the solutions to equation (12) with $1\le a < x \le 100$ are
$x=18,~a=1,m=3,y=7;~a=7,m=4,y=7;~a=8,m=3,y=14$;~$x=22,~a=3,m=2,y=39;
~a=12, m=2, y=78$; $x=30,~a=19, m=2,y=133$ and $x=68,~a=13,m=2,y=247;~
a=52,m=2,y=494$, and
when $n=4$, the solutions are $x=7,~a=1,m=2,y=20; ~a=4,m=2,y=40$; 
$x=41,~a=21,m=2,y=1218$; $x=99,~a=58,m=2,y=7540$ [26]. 

If $n~=~\prod_j~{\tilde p}_j^{t_j}$, 
$$\eqalign{{{x^n~-~1}\over {x~-~1}}~&=~\prod_{k=1}^{n-1}~
[x~-~e^{{2\pi k i}\over n}]
\cr
~&=~
{{x^{{\tilde p}_j^{t_j}}~-~1}\over {x~-~1}}~\cdot~
(x^{n-{\tilde p}_j^{t_j}}~+~x^{n-2{\tilde p}_j^{t_j}}~+~...~+~
x^{{\tilde p}_j^{t_j}}~+~1)
\cr}
\eqno(17)
$$
and ${{x^{{\tilde p}_j^{t_j}}~-~1}\over {x~-~1}}$ is a factor of ${{x^n~-~1}
\over {x~-~1}}$.

For every integer $k_1$ between $1$ and $n-1$, the product 
$[x~-~e^{{2\pi i {k_1}}\over n}][x~-~e^{{2\pi i {k_2}}\over n}]$
is a real quadratic polynomial
$$x^2~-~2~cos~\left({{2\pi {k_1}}\over n}\right)x~+~1
\eqno(18)
$$
when $k_2~=~n~-~k_1$, and the coefficient of $x$ is an integer when 
$cos~\left({{2\pi{k_1}}\over n}\right)~=~\pm {1\over 2}$, which implies 
that $k_1~=~{n\over 6},~{n\over 3},~{{2n}\over 3},~{{5n}\over 6}$, providing 
two quadratic factors with integer coefficients, when $n$ is divisible 
by $3$, and four quadratic factors with integer coefficients, when $n$ is 
divisible by $6$.  The trinomial $x^2+x+1$ is a particular example of 
${{x^{{\tilde p}_j^{t_j}} - 1}\over {x-1}}$, and the higher-degree polynomial 
also could be examined for its primality if tests were available. 

When $x$ is a prime $p$, the quadratic factor (18) is an integer
$p^2~\mp~k~+~1$ if
\hfil\break
$cos~\left({{2\pi {k_1}}\over n}\right)~=~\pm {k\over {2p}}$.
Primality tests for trinomials of the type $Ax^2~+~Bx~-~1$ have been developed
[27]. They can be adapted to the present case by setting $A~=~1$,
$B~=~2{\sqrt{ k + 2}}$, $x~=~p-{\sqrt{ k~+~2}}$.  Further restricting
$k$ to be $\kappa^2~-~2$ for some integer $\kappa$, $x$ is then an integer
which can be factorized as $x~=~ar$, where $r$ is a fixed prime.
Suppose that $r > a^2~+~2\kappa a$.  If 
some prime factor of $x^2+2\kappa x-1$ is equal to $\pm 1~(mod~r)$,
then the trinomial is also a prime number [27].   Similarly, if $r$ is an
odd prime, 
$r~\ge~{{(a^2~-~3)}\over 2}~+~2\kappa a$, and when $2\vert a$, 
$r > {{({{a^2}\over 2}+2 \kappa)}\over 8}$, then the trinomial $x^2~+~
2\kappa x~-~1$ is prime [27].

Since the solution to $x^2~+~x~+~1~=~{{y^2}\over a}$ is
$$x~=~{{-1~\pm~{\sqrt{{{4y^2}\over a}~-~3}}}\over 2}
\eqno(19)
$$
it will be integer only if
$$y~=~{\sqrt{a(z^2~+~3)} \over 2}~~~,~~~ z~\in~{\Bbb Z} 
\eqno(20)
$$
is integer.  If $z~>~1$, $(z~+~1)^2~-~z^2~=~2z~+~1~>~3$ and ${\sqrt{z^2~+~3}}$
is not rational, which confirms that there are no integer solutions to
the original equation when $a=1$, except when $x=0$ or $x=-1$.  Integer 
solutions to equation (21) are determined by solutions of the quadratic 
Diophantine equation 
$$z^2~-~D~r^2~=~-3
\eqno(21)
$$
This equation has been investigated using the continued fraction expansion
of ${\sqrt{D}}$ [28].
Ordering the integer solutions of this equation by the magnitude of
$z+r{\sqrt D}$, the fundamental solution, given by the smallest value
of this expression, shall be denoted by the pair of integers $(z_1,~r_1)$.
For any solution $(x,y)$ of the Pell equation $x^2~-~Dy^2~=~1$, an
infinite number of solutions of equation (22) are generated by the
identity
$$(z_1~+~r_1{\sqrt D})(x~+~y{\sqrt D})~=~z_1x~+~r_1y D~+~(z_1y~+~r_1x)
~{\sqrt D}
\eqno(22)
$$
as the pairs of integers 
$\{(z_1x~+~r_1y D,~z_1 y~+~r_1 x)~\vert~x^2~-~Dy^2~=~1\}$ define a class
of solutions to equation (21). Except when $D~\equiv~1~(mod~4)$, if $D$ is 
not a perfect square and is a multiple of 3, then there is only one class of 
solutions, whereas, if $D$ is 
not a multiple of 3, there are two classes of solutions.  If
$D~\equiv~1~(mod~4)$, there may be one class or two classes of solutions [29].

The solutions (22) usually will be pairs of composite integers.  Although
only prime values of $x$ are included in equation (11), $z=2x+1$ 
may be a composite number belonging to a large set of similar solutions,
or $z=2x+1$ is a prime belonging to a restricted set of solutions. 
The density of primes $x~\le~T,~T~\gg~1$ producing prime values of $z$ 
satisfying (22) tends to ${2\over {(T~-~1)~ln~T}}$ as $T~\to~\infty$.  

Given any two solutions to equation (22), $(z_1,~r_1)$ and $(z_2,~r_2)$,
it follows that
$${{x_1^3~-~1}\over {x_1~-~1}}~{{x_2^3~-~1}\over {x_2~-~1}}~=~
{{(z_1^2~+~3)}\over 4}\cdot {{(z_2^2~+~3)}\over 4}~=~
\left({{D~r_1^2}\over 4}\right)\cdot \left({{D~r_2^2}\over 4}\right)~=
~\left({{D~r_1~r_2}\over 4}\right)^2
\eqno(23)
$$ 

Even though the quotients ${{x^3~-~1}\over {x~-~1}}$ are not perfect squares in
general, the products of such quotients can be perfect squares.  This
property implies that there are potential cancellations and combinations
of factors in the expression (11), and a more detailed study
of the divisors of the repunits ${{x^n~-~1}\over {x~-~1}}$ is required to
determine if the product in (11) is the square of a rational number.

The repunits in equation (11) generally will not be equal because the
only known integer solutions [30][31][32] to the equation
$${{x^m~-~1}\over {x~-~1}}~=~{{y^n~-~1}\over {y~-~1}}
~~~x~\ne~y,~m~\ne~n
\eqno(24)
$$
are
$$\eqalign{31~&=~{2^5~-~1}~=~{{5^3~-~1}\over {5~-~1}}
\cr
8191~&=~{2^{13}~-~1}~=~{{90^3~-~1}\over {90~-~1}}
\cr}
\eqno(25)
$$
It has been conjectured that there are only finitely many solutions to (24)
and the following result has been proven [30].

\noindent
{\bf Theorem} (Shorey 1989) - Let $N~>~2$, $N~\ne~31$ and $N~\ne~8191$
be an integer and assume that the number of distinct prime factors of
$N~-~1$ is less than or equal to $5$.  There is at most one element
$y\in S(N)$, the set of all integers $x$, $1~<~x~<~N~-~1$, such that $N$
has all of the digits equal in its $x$-adic expansion, where the
integer $l(N;y)$, defined by
$$N~-~1~=~y~{{{y^{l(N;y)~-~1}~-~1}}\over {y~-~1}}
\eqno(26)
$$
is odd.

The repunit ${{x^n~-~1}\over {x~-~1}}$ is the Lucas sequence
$$U_n(a,b)~=~{{\alpha^n~-~\beta^n}\over {\alpha~-~\beta}}
\eqno(27)
$$
with $\alpha~=~x$, $\beta~=~1$, derived from the second-order recurrence  
$$U_{n+2}(a,b)~=~a~U_{n+1}(a,b)~-~b~U_n(a,b)
\eqno(28)
$$
with parameters $a~=~\alpha~+~\beta~=~x~+~1$ and $b~=~\alpha~\beta~=~x$.
For a primary recurrence, defined by the initial values $U_0~=~0$ and
$U_1~=~1$, the rank of apparition of a prime $p$ is the least positive
integer $k$, if it exists, such that $U_k~\equiv~0~(mod~p)$ [33].  
When $b~\ne~0$, the values of the rank of apparition, denoted by 
$\alpha(a,b,p)$ are listed.
$$\eqalign{\alpha(x~+~1,x,p)~&=~ord_p(x)
\cr
if~(x/p)~=~-1~and~p~\equiv~3~(mod~4),~then~\alpha(x~+~1,x,p)~&\equiv~2~(mod~4)
\cr
if~(x/p)~=~-1~and~p~\equiv~1~(mod~4),~then~\alpha(x~+~1,x,p)~&\equiv~0~(mod~4)
\cr
if~(x/p)~=~1~and~p~\equiv~3~(mod~4),~then~\alpha(x~+~1,x,p)~&\equiv~1~(mod~4)
\cr
if~(x/p)~=~1~and~p~\equiv~1~(mod~4),~then~\alpha(x~+~1,x,p)~&\equiv~0~or~2~
(mod~4)
\cr}
\eqno(29)
$$
The numerical values of $ord_p(x)$ have been tabulated, but the functional 
dependence of $\alpha(x+1,x,p)$ would be based on rules for $ord_p(x)$, 
which have yet to be formulated.  The extent to which the arguments $a$ and 
$b$ determine the divisibility of $U_n(a,b)$ [6] can be summarized as follows:

\noindent
Let $p$ be an odd prime.
\hfil\break
If $p\vert a$, $p\vert b$, then $p\vert U_n(a,b)$ for all $n > 1$.
\hfil\break
If $p\vert a$ and $p\not\vert~b$, then $p\vert U_n(a,b)$ exactly when $n$
is even.
\hfil\break
If $p\not\vert~a$, $p\not\vert~b$, $p\vert D=a^2-4b$, then $p\vert U_n(a,b)$
when $p\vert n$.
\hfil\break
If $p\not\vert~abD$, then $p\vert U_{p-(D/p)}(a,b)$.

The possibility of products of sums of powers of primes being perfect
squares can be ascertained from the listing of the square classes of Lucas
sequences.  It is known that the square classes \footnote{*}{$\square$
represents the square of a rational number} 
of $U_n(a,b)$ [34], defined by
$$\eqalign{x^2 U_m(a,b)~&=~y^2 U_n(a,b)
\cr
U_m(a,b)~U_n(a,b)~&=~\square
\cr}
\eqno(30)
$$
are 
$$\eqalign{\{U_1,U_2,U_3\}~&or~\{U_5,U_{10}\}~when~a~\equiv~b~\equiv~1~(mod~4)
\cr
\{U_1, U_3, U_5\}~&when~b~\equiv~1~(mod~4)~and~a~\equiv~3~(mod~4)
\cr
\{U_1,U_2,U_{12}\}~&~or~\{U_3,U_6\}~when~b~\equiv~3~(mod~4)
\cr
\{U_m,U_{3m}\},~&~m > 1,~m~odd,~3~\not\vert~m,~a~<\vert b+1\vert,~
b~\equiv~1~(mod~4)
\cr
~&(-b\vert a)=1~for~at~most~a~finite~set~of~values~
of~m
\cr}
\eqno(31)
$$ 
although they consist of only one element when
$U_n(a,b)~=~{{x^n~-~1}\over {x~-~1}}$ [35].  Additionally, through the 
application of Jacobi symbols to Lucas numbers, it has been demonstrated 
that
$${{\alpha^n~-~\beta^n}\over {\alpha~-~\beta}}~\ne~n\square
\eqno(32)
$$
when $gcd(\alpha,\beta)~=~1$, $\alpha\beta~\equiv~0$ or $3~(mod~4)$ and
$n$ is an odd integer greater than one [36].

Given the prime decomposition of $n~=~{\tilde p}_1~...~{\tilde p}_r$, the 
product of cyclotomic polynomials 
$${{x^n~-~1}\over {x~-~1}}~=~\Phi_{{\tilde p}_1}(x)~\Phi_{{\tilde p}_2}
(x^{{\tilde p}_1})~\Phi_{{\tilde p}_3}(x^{{\tilde p}_1{\tilde p}_2})~...~
\Phi_{{\tilde p}_r}(x^{{\tilde p}_1 {\tilde p}_2 ... {\tilde p}_{r-1}})
\eqno(33)
$$  
provides a factorization of the quotient ${{q_i^{2\alpha_i+1}~-~1}\over 
{q_i~-~1}}$.   Even though $\Phi_n(x)$ is irreducible over ${\Bbb Q}$,
none of the factors above are necessarily prime when evaluated 
at $q_i$.  The number $\Phi_p(q)~=~{{q^p~-~1}\over {q~-~1}}$ arises in the
study of the solvability of groups of odd order [37] and its
composite nature is revealed in computations of the greatest common divisor
of ${{p^q~-~1}\over {p~-~1}}$ and ${{q^p~-~1}\over {q~-~1}}$ [38].
Only a single example of the relation
$$q^{\prime a}~=~{{q^p~-~1}\over {q~-~1}}~~~~~~~~~~p,~q,~q^\prime~prime
\eqno(34)
$$
is known for primes, and there exists a finite bound for the number of 
solutions [39].

Nevertheless, conditions on prime divisors of Lucas sequences provide
information on the  factorization of cyclotomic polynomials in 
equation (33).  Lower bounds for the greatest prime divisors of 
Lucas numbers have been given in a series of articles 
[40]-[46].
Specifically, if $P(k)$ denotes the largest prime divisor of $k$,
then
$$P\left({{\alpha^n~-~\beta^n}\over {\alpha~-~\beta}}\right)
~\ge~P(\Phi_n(a,b))~>~C~{{n~log~n^{1-\kappa~log~2}}\over {log~log~log~n}}
\eqno(35)
$$
Refined estimates of the greatest prime divisor have yet to be obtained,
but the value of the largest primitive divisor, which divides $U_n(a,b)$ but
not $U_m(a,b)$ for $m~<~n$, can be deduced.
Since $x^n~-~1~=~\prod_{d\vert n}~\Phi_d (x)$ where $\Phi_n(x)$ is the
$n^{th}$ cyclotomic polynomial, it can be shown that the largest primitive
factor
[41][47][48] of $x^n~-~1$ when $x~\ge~2$ and $n~\ge~3$ is
$$\eqalign{\Phi_n(x)&~~~~~~~~~~if~\Phi_n(x)~and~n~are~relatively~prime
           \cr
           {{\Phi_n(x)}\over p}&~~~~~~~~if~a~common~prime~factor~p~of
                                       ~\Phi_n(x)~and~n~exist
                                                       \cr}
\eqno(36)
$$
In the latter case, if $n~=~p^\alpha p^{\prime \alpha^\prime}
p^{\prime\prime \alpha^{\prime\prime}}~...$ is the prime factorization
of $n$, then $\Phi_n(x)$ is divisible by $p$ if and only if
$h~=~{n\over {p^\alpha}}$ is the multiplicative order of $x$ modulo $p$.
Moreover $p\vert\vert \Phi_{hp^j},~j~\ge~0$ when $p$ is an odd prime.
The prime factors of $\Phi_n(q)$ either satisfy $p~=~nk+~1$ or 
$p\vert n$.

Selecting the largest primitive factors of each repunit 
${{q_i^{n_i}~-~1}\over {q_i~-~1}}$ in the expression (11), the product of 
these factors when $q_i-1\not\vert~\Phi_{n_i}(q_i)$ takes the form
$${{\Phi_{n_1}(q_1)}\over {p_1}}~{{\Phi_{n_2}(q_2)}\over {p_2}}~...~
{{\Phi_{n_l}(q_l)}\over {p_l}}~
\times~{1\over 2}~\left[{{\Phi_{4m+2}(4k+1)}\over {p_{l+1}}}\right]^{-1}
\eqno(37)
$$ 
where the indices are odd numbers $n_i~=~2\alpha_i~+~1$, $p_i,~i~=~1,...,l$, 
represents the common factor of $n_i$ and $\Phi_{n_i}(q_i)$, and
$p_{l+1}$ is a common factor of $4m~+~2$ and $\Phi_{4m+2}(4k+1)$.
If $q_i-1 \vert \Phi_{n_i}(q_i)$, then it should be included in the 
denominator with $p_i$, so that the relevant factor in (37) 
becomes ${{\Phi_{n_i}(q_i)}\over {(q_i-1)p_i}}$.

Cyclotomic polynomials are known to have the following properties:

\noindent
(i) $\Phi_n(x)$ are strictly increasing functions for $x~\ge~1$ [49], so that
$\Phi_n(q_j)~>~\Phi_n(q_i)$ when 
\hfil\break
\phantom{.....}$q_j$ is the larger prime.
\vskip 5pt
\noindent
(ii) If $n~>~1$ is square-free, then 
$$\Phi_n(x)~=~{{\Phi_{n\over p}(x^p)}\over {\Phi_{n\over p}(x)}}
\eqno(38)
$$
\phantom{.....} when $p$ is a prime factor of $n$ [50].

From the first property, it follows that equality of $\Phi_{n_i}(q_i)$
and $\Phi_{n_j}(q_j)$ could only be achieved if $n_i~\ne~n_j$.
The factor $p_i$ is equal to the greatest prime divisor of $n_i$ which
also divides $\Phi_{n_i}(q_i)$ if $gcd(n_i,\Phi_{n_i}(q_i))\ne 1$ [51].
The ratio of ${{\Phi_{n_i}(q_i)}\over {p_i}}$ and ${{\Phi_{n_j}(q_j)}
\over {p_j}}$.  
Since $n_i~\ne~n_j$ and not all of their prime factors are equal, the ratio
${{\Phi_{n_i}(q_i)}\over {\Phi_{n_j}(q_j)}}$ will contain fractions
of the type ${{\Phi_{{\tilde p}_i}(q_i^{{\tilde t}_i})}\over
{\Phi_{{\tilde p}_j}(q_j^{{\tilde t}_j})}}$ where ${\tilde p}_i$ and 
${\tilde p}_j$ are prime factors not common to both $n_i$ and $n_j$.  A 
study of the values of the cyclotomic polynomial would be necessary 
to establish that these ratios contain irreducible fractions.

There are at least eight prime factors of an odd perfect number [52], and
when it is relatively prime to 3, there is a minimum of eleven prime factors
[52][53].
More recently, a proof has been given that there must be at least 
14 distinct prime factors in any odd perfect number and the sum of the
powers of these prime factors must be greater than or equal to 29 [54]. 
By the previous argument, the cyclotomic polynomials evaluated at different 
prime values could only be equal if the indices $n_i$ are
distinct.  If the power $n_i~=~2\alpha_i~+~1$ are all distinct,
then at least one of the powers will be greater than 30 if there are more 
than 14 different prime factors and then the repunit 
${{q_i^{n_i}~-~1}\over {q_i~-~1}}$ would have a primitive divisor when the 
logarithmic height $h({\beta\over \alpha})$ is less than or
equal to $4$ [55].  There are indications that a primitive divisor will 
exist generally when $n_i > 30$ [56].
If it can be demonstrated that the number of repunits in
the product (11) is greater than 14, then by the conjecture on primitive
divisors of Lucas sequences, it follows that at least one of the repunits will
have a primitive divisor that is not matched elsewhere.

\vskip 30pt
\centerline{\bf Acknowledgements}
\noindent
I would like to acknowledge useful discussions with Prof. G. L. Cohen, Dr
R. Melham, 
\hfil\break
Dr A. Nelson, Mr A. Steel and Prof. A. J. van der Poorten.

\vfill\eject
\noindent
\centerline{\bf References}
\parskip=5pt
\item{[1]} Euclid, ${\underline{Elements}}$ Book IX, Proposition 36,
Opera, 2 (Leipzig, 1884) p. 408
\item{[2]}  M. Crubellier and J. Sip,~`Looking for Perfect Numbers', Chap. 15, 
in ${\underline{History~of}}$
\hfil\break
${\underline{~Mathematics~-~Histories~of~Problems}}$,
The Inter-IREM Commission: Epistemology and History of Mathematics, translated
by Chris Weeks (Paris: Ellipses, 1997) 389 - 410
\item{[3]} E. Lucas, Assoc. Francaise p. l'Avanc. des Sciences,
${\underline{5}}$ (1876) 61 - 68 
\item{[4]} D. H. Lehmer, J. London Math. Soc. ${\underline{10}}$ (1935)
162 - 165
\item{[5]} P. de Fermat, letter to Mersenne, 1640, in Ouvres de Fermat,
2 (Paris, 1894) pp. 198 - 199
\item{[6]} P. Ribenboim, ${\underline{The~New~Book~of~Prime~Number~Records}}$
(New York: 
\hfil\break
Springer-Verlag, 1996)
\item{[7]}  L. Euler, Nuov. Mem. Acad. Berlin in 1772, p. 35
\item{[8]} D. Shanks, ${\underline{Solved~and~Unsolved~Problems~in~
Number~Theory}}$ (Washington: Spartan Books, 1962)
\item{[9]} J. Brillhart, D. H. Lehmer and J. L. Selfridge,
Math. Comp. ${\underline{30}}$ (1975) 620 - 647
\item{[10]} {\"O}. J. R{\"o}dseth, BIT ${\underline{34}}$ (1994)
451 - 454
\item{[11]} M. R. Heyworth, New Zealand Math. Mag. ${\underline{19}}$
(1982/83) no. 2, 63 - 69
\item{[12]} J. R. Clay, Y.-N. Yeh, Periodica Mathematica Hungarica, Vol. 29(2)
(1994) 137 - 157
\item{[13]} B. de la Rosa, Fibonacci Quart., Vol. 16, No. 6 (1978) 518 - 522
\item{[14]} R. K. Guy, Fibonacci Quart., Vol. 20, No. 1 (1982) 36 - 38

\item{[15]} A. G. Wolman, J. Reine Angew. Math. ${\underline{477}}$ (1996)
31 - 70
\item{[16]} M. Atiyah, Ann. Scient. Ecole Norm. Sup. V ${\underline{4}}$
(1971) 47 - 62
\item{[17]} D. Johnson, J. London Math. Soc. ${\underline{22}}$ (1980)
365 - 373
\item{[18]} S. Feigelstock, Math. Mag. ${\underline{49}}$(4)(1976) 198 - 199
\item{[19]} R. P. Brent, G. L. Cohen and H.J.J te Riele,
Math. Comp. ${\underline{57}}$ (1991) 857 - 868
\item{[20]} P. Hagis and G. L. Cohen, Math. Comp. ${\underline{67}}$
(1998) 1323 - 1330
\item{[21]} M. Kishore, Math. Comp. ${\underline{32}}$ (1978) 303 - 309
\item{[22]}  L. Euler, De Numeris Amicabilus in 
${\underline{Opera~Omnia}}$, I,2 (Leipzig, 1944) pp. 353 - 365
\hfil\break
J. A. Ewell, Journal of Number Theory ${\underline{12}}$ (1980) 339 - 342
\item{[23]} T. Nagell, Norsk. Mat. Tidsskr. ${\underline{2}}$ (1920)
75 - 78
\hfil\break
T. Nagell, Mat. Fornings Skr. ${\underline{1}}$(3)(1921) 
\item{[24]} W. Ljunggren, Norsk Mat. Tidsskr. ${\underline{25}}$ (1943) 17 - 20
\item{[25]} P. Ribenboim, ${\underline{Catalan's~Conjecture: 
~Are~8~and~9~the~Only~Consecutive~Powers?'}}$ (Sydney: Academic Press, 1994)
\item{[26]} K. Inkeri, Acta Arithmetica, Vol. XLVI (1972) 299 - 311
\item{[27]} H. C. Williams, Pac. Journal of Math. ${\underline{98}}$(2)
(1982) 477 - 494
\item{[28]} R. A. Mollin, A. J. van der Poorten and H. C. Williams,
Journal de Theorie des Nombres Bordeaux ${\underline{6}}$ (1994) 421 - 459
\item{[29]} T. Nagell, ${\underline{Introduction~to~Number~Theory}}$
(New York: Chelsea Pub. Co., 1964)
\item{[30]} J. H. Loxton, Acta Arithmetica, Vol. XLVI (1986) 113 - 123
\item{[31]} T. N. Shorey, Acta Arithmetica, Vol. LIII (1989) 187 - 205
\item{[32]} Yu. V. Nesterenko and T. N. Shorey, Acta Arith.
${\underline{83}}$ (1998) 381 - 389
\item{[33]} L. Somers, Fib. Quart. ${\underline{18}}$(4) (1980) 316 - 334
\item{[34]} P. Ribenboim and W. L. McDaniel, C. R. Math.
Rep. Acad. Sci. Canada, Vol. XVIII, No. 5 (1996) 223 - 227
\item{[35]} P. Ribenboim, J. Sichuan Univ., Vol. 26, Special Issue (1989) 
                                                                196 - 199 
\item{[36]} A. Rotkiewicz, Acta Arithmetica, Vol. XLII (1983) 163 - 187
\item{[37]} W. Feit and J. C. Thompson, Proc. Nat. Acad. Sci. U.S.A. 
${\underline{48}}$ (1962) 968 - 970
\item{[38]} N. M. Stephens, Math. Comp. ${\underline{25}}$ (1971) 625
\item{[39]} D. Estes, R. Guralnick, M. Schacher and E. Straus, Pac. Journal 
of Math. ${\underline{118}}$ (1985) 359 - 367
\item{[40]} K. Zsigmondy, Monatsh. fur Math. ${\underline{3}}$ (1892) 
265 - 284
\item{[41]} G. D. Birkhoff and H. S. Vandiver, Annals of Math.,
${\underline{5}}$ (1904) 173 - 180
\item{[42]} R. D. Carmichael, Ann. of Math. ${\underline{15}}$ (1913) 30 - 70
\item{[43]} A. Schinzel, Proc. Cambridge Phil. Soc. ${\underline{58}}$ 
(1962) 555 - 562
\item{[44]} C. L. Stewart, Acta Arith.  ${\underline{26}}$ (1975) 
427 - 433; J. Reine Angew. Math. ${\underline{333}}$ (1982) 12 -31
\item{[45]} K. Gy{\"o}ry, Acta Arith. ${\underline{40}}$ (1981/82) 
369 - 373
\item{[46]} P. Kiss,`On primitive prime power divisors of Lucas numbers'
in ${\underline{Number~theory}}$, Vol. II, Colloq. Math. Soc. J{\'anos} 
Bolyai, 51 (Amsterdam: North-Holland Publishing Co., 1990)
\item{[47]} L. E. Dickson, Amer. Math. Monthly ${\underline{16}}$ (1905)
86 - 89
\item{[48]} R. A. Rueppel and O. J. Staffelbach, IEEE Transaction on
Information Theory, Vol. IT-33, No. 1 (1987) 124 - 131
\item{[49]} K. Motose, Math. Journal of Okayama Univ. ${\underline{35}}$
(1993) 35 - 40; Math. Journal of Okayama Univ. ${\underline{37}}$ (1995)
27 - 36
\item{[50]} R. P. Brent, Math. Comp. ${\underline{61}}$ (1993) 131 - 149
\item{[51]} S. Goulomb, Amer. Math. Monthly ${\underline{85}}$ (1978) 734 - 737
\item{[54]} P. Hagis, Math. Comp. ${\underline{35}}$ (1980) 1027 - 1032;
Math. Comp. ${\underline{40}}$ (1983) 399 - 404
\item{[53]} M. Kishore, Math. Comp. ${\underline{40}}$ (1983) 405 - 411
\item{[54]} M. D. Sayers, ${\underline{An~improved~lower~bound~for~the~
total~number~of~prime~factors~of}}$
\hfil\break
${\underline{an~odd~perfect~number}}$, M. App. Sc. Thesis -
New South Wales Institute of Technology, 1986
\item{[55]}  P. M. Voutier, Math. Comp. ${\underline{64}}$ (1995) 869 - 888; 
Journal de Theorie des Nombres de Bordeaux ${\underline{8}}$ (1996) 251 - 274
\item{[56]}  P. M. Voutier, Math. Proc. Cambridge 
Philos. Soc. ${\underline{123}}$ (1998) 407 - 419

\end